\documentclass[aps,prb,twocolumn,superscriptaddress,showpacs]{revtex4-1}

\usepackage{graphicx}
\usepackage{calc}
\usepackage{bm}
\usepackage{color}
\usepackage{mathtools}

\begin{document}

\title{Crossover from relativistic to non-relativistic net magnetization for MnTe altermagnet candidate}

\author{N.N. Orlova}
\author{A.A. Avakyants}
\author{A.V.~Timonina}
\author{N.N.~Kolesnikov}
\author{E.V.~Deviatov}
\affiliation{Institute of Solid State Physics of the Russian Academy of Sciences, Chernogolovka, Moscow District, 2 Academician Ossipyan str., 142432 Russia}

\date{\today}

\begin{abstract}
We experimentally study magnetization reversal curves for MnTe single crystals, which is the altermagnetic candidate. Above 85~K temperature, we confirm the antiferromagnetic  behavior of magnetization $M$, which is known for $\alpha$--MnTe. Below 85~K, we observe anomalous low-field magnetization behavior, which is accompanied by the sophisticated  $M(\alpha)$ angle dependence with beating pattern as the  interplay between $M(\alpha)$ maxima and minima: in low fields,   $M(\alpha)$ shows ferromagnetic-like 180$^\circ$ periodicity, while at high magnetic fields, the periodicity is changed  to the 90$^\circ$ one. This angle dependence is the most striking result of our experiment, while it can not be expected for standard magnetic systems. In contrast, in altermagnets,  symmetry allows ferromagnetic behavior only due to the  spin-orbit coupling. Thus, we claim that our experiment shows the effect of weak spin-orbit coupling in MnTe, with crossover from relativistic to non-relativistic net magnetization, and, therefore, we experimentally confirm altermagnetism in MnTe. 
\end{abstract}

\maketitle

\section{Introduction}

Recent interest to topological materials is mostly connected with topological surface states, which appear due to the bulk-boundary correspondence: inversion of the  bulk spectrum leads to the gapless surface states at the interface,  even if the the gap is still present in the bulk~\cite{Volkov-Pankratov}. Topological surface states are always characterized by spin-momentum locking~\cite{Volkov-Pankratov,MZHasan,Armitage,Fermi arc}, so one can expect complicated spin textures even for nonmagnetic materials with strong spin-orbit coupling (SOC).  In Weyl semimetals~\cite{Armitage}, spin is rotating along the Fermi-arc~\cite{Fermi arc-SOC}. The spin structure is more complicated for helical surface states in Dirac semimetals~\cite{Armitage} and topological insulators~\cite{MZHasan}. In the topological nodal-line semimetals~\cite{Armitage}, the drumhead surface states lead to the spin textures of the skirmion type~\cite{nodal-line}. 

Topological surface states can also appear in magnetic materials~\cite{mag1,mag2,mag3}. Breaking of the time-reversal symmetry  is responsible for the  Weyl nodes separation in the k-space for magnetic Weyl semimetals~\cite{Armitage}, and, therefore, for  bulk  and surface spin textures  due to the spin-momentum locking~\cite{spin1,spin2,spin3}.

Recently, the concept of spin-momentum locking was extended to the case of weak spin-orbit coupling, i.e. to the non-relativistic groups of magnetic symmetry~\cite{alter_common,alter_mazin}. 
As a result, a new class of altermagnetic materials has been added to  usual ferro- and antiferromagnetics. Previously, this class of materials has not been considered both in usual magnetic classes without spin-momentum locking, and in the topological relativistic magnetic symmetry groups with strong spin-momentum locking. For altermagnetics, the small net magnetization is accompanied by alternating spin-momentum locking in the k-space, so the unusual spin splitting is predicted~\cite{alter_common,alter_josephson}. For example, RuO$_2$ altermagnet consists of two spin sublattices with orthogonal spin directions~\cite{AHE_RuO2}. In the k-space, the up-polarized subband can be obtained by   $\pi/2$ rotation of the down-polarized subband, so RuO$_2$ altermagnet is characterized by d-wave order parameter~\cite{alter_supercond_notes,alter_normal_junction}. The  probability to scatter between subbands depends both on the electron spin and the propagation direction due to the spin-momentum locking~\cite{AHE_MnTe1}.    

As a result, altermagnetics are characterized by sophisticated spin structures, which should lead to different physical phenomena. For example, anomalous Hall effect (AHE) is predicted for altermagnetics~\cite{alter_original}, despite of the zero nonrelativistic net magnetization~\cite{alter_mazin}. AHE has been experimentally demonstrated~\cite{AHE_RuO2,AHE_MnTe1,AHE_MnTe2,AHE_Mn5Si3} for some altermagnetic candidates, MnTe, Mn$_5$Si$_3$ and RuO$_2$. In contrast to RuO2~\cite{AHE_RuO2}, the measurements in MnTe and Mn$_5$Si$_3$ show hysteresis and spontaneous AHE signals at zero magnetic field~\cite{AHE_MnTe1,AHE_MnTe2,AHE_Mn5Si3}.

MnTe  is an intrinsic room-temperature magnetic semiconductor with a collinear antiparallel magnetic ordering of Mn moments~\cite{MnTe1,MnTe2,MnTe3,MnTe4,MnTe5,MnTe6}.  Recently it was argued,  that the spontaneous nature of the AHE still requires relativistic spin-orbit interaction~\cite{AHE_MnTe1,AHE_MnTe2}. Despite the expected vanishing net magnetization in altermagnetics~\cite{alter_mazin}, the bulk form of MnTe exhibits small but detectable magnetic moment correlating with hysteretic behavior of the AHE~\cite{AHE_MnTe2}. In the same time, the AHE signal does not correlate with the angle-dependence of the weak saturation magnetization for Mn$_5$Si$_3$, which is isotropic~\cite{AHE_Mn5Si3}. Inconsistency between the expected zero non-relativistic net magnetization and ambiguous experimental magnetization behavior  requires comprehensive magnetization measurements in altermagnetics in wide temperature and magnetic field ranges. This investigation can be conveniently performed for MnTe, which has been studied both experimentally~\cite{MnTe1,MnTe2,MnTe3,MnTe4,MnTe5,MnTe6} and theoretically~\cite{MnTe_Mazin}. MnTe is also characterized by accessible (2--3~T) magnetic field range~\cite{AHE_MnTe1,AHE_MnTe2,AHE_Mn5Si3} in contrast to RuO2~\cite{AHE_RuO2} altermagnetic candidate.

Here,  we experimentally study magnetization reversal curves for MnTe single crystals, which is the altermagnetic candidate. Above 85~K temperature, we confirm the antiferromagnetic  behavior of magnetization $M$, which is known for $\alpha$--MnTe with 307--325~K N\'eel temperature. Below 85~K, we observe anomalous low-field magnetization behavior, which is accompanied by the sophisticated  $M(\alpha)$ angle dependence with beating pattern as the  interplay between $M(\alpha)$ maxima and minima.

\section{Samples and technique}

\begin{figure}
\includegraphics[width=1\columnwidth]{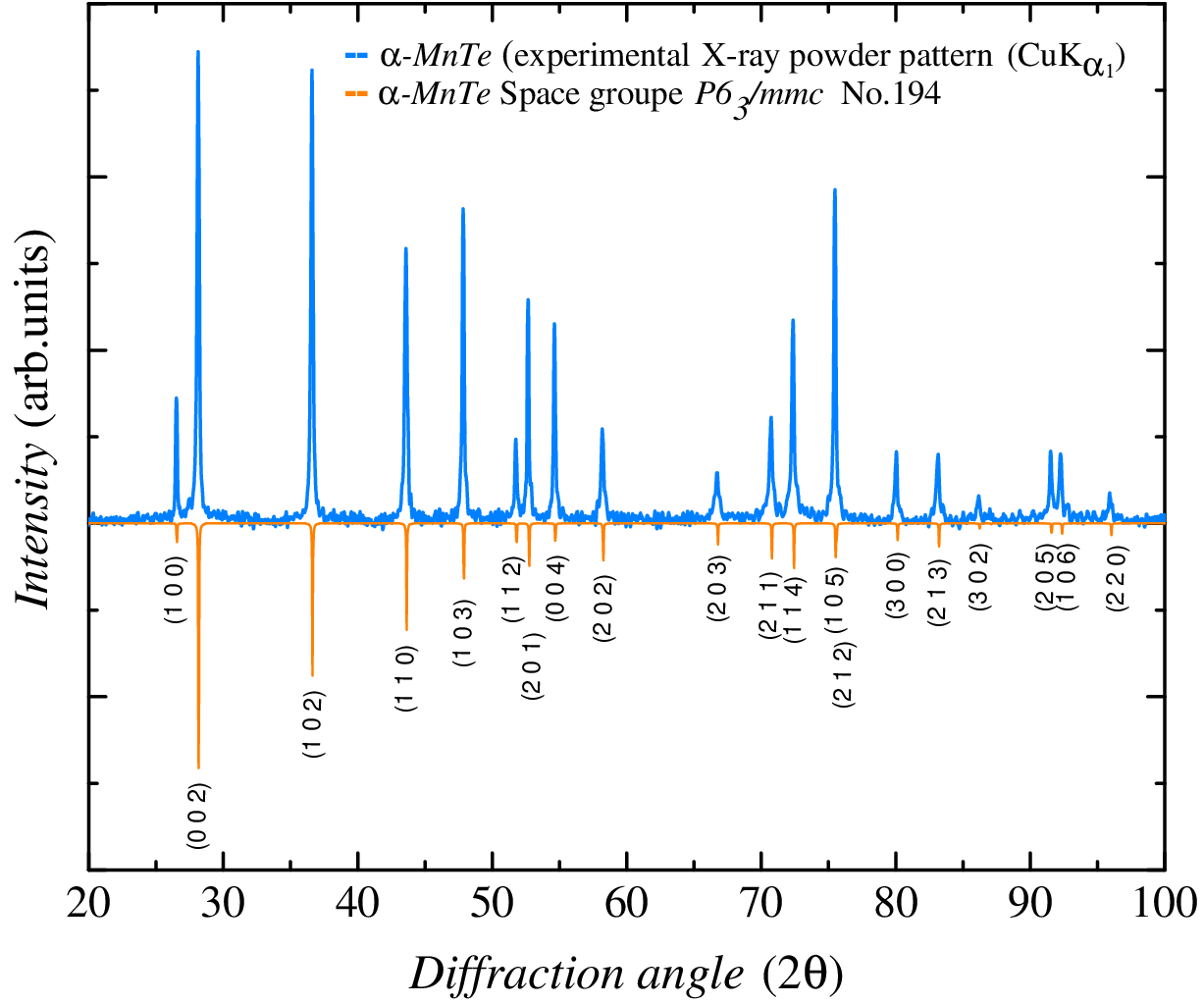}
\caption{(Color online)  The X-ray powder diffraction  pattern (Cu K$_{\alpha1}$ radiation), which is obtained for the crushed MnTe single crystal. The single-phase  $\alpha$-MnTe is confirmed with the space group $P6_3 /mmc$ No. 194, so the results below cannot appear from the incorrect stoichiometry or oxides~\cite{impur1,impur2}.  
  }
\label{sample}
\end{figure}

For investigation of low magnetic moment, we use MnTe single crystals  which is preferable in comparison with thin films to avoid admixture of the substantial signal from the substrate.

MnTe was synthesized by reaction of elements (99.99\% Mn  and 99.9999\% Te) in evacuated silica ampules slowly heated up to 1050--1070$^\circ$C. The obtained loads were melted in the graphite crucibles under 10 MPa argon pressure, then homogenized at 1200$^\circ$C for 1 hour. The crystals grown by gradient freezing method are groups of single crystal domains with volume up to 0.5--1.0~cm$^3$. The MnTe composition is verified by energy-dispersive X-ray spectroscopy. The powder X-ray diffraction analysis confirms single-phase $\alpha$-MnTe with the space group $P6_3 /mmc$ No. 194, see Fig.~\ref{sample}. 

A small (0.53~mg -- 5.04~mg) mechanically cleaved single crystal sample, see the inset to Fig.~\ref{fig2}, is mounted to the sample holder by the low temperature grease, which has small diamagnetic response below 200~K. We check, that without MnTe sample we obtain strictly linear diamagnetic dependence, with -100~$\mu$emu value in 10~kOe for the same amount of grease as for the sample in Fig.~\ref{fig2}.  For the smaller samples the amount of grease is also diminished, so we can estimate the diamagnetic contribution (sample holder and grease) as about 10\% of the measured value. This contribution is verified to be strictly linear, thus, the experimental setup allows high-resolution measurements in low fields. For these reasons, and to provide direct comparison between the data, we do not subtract this line from the presented $M(H)$ curves.
     
To investigate magnetic properties, we use Lake Shore Cryotronics 8604 VSM magnetometer, equipped with nitrogen flow cryostat.  We investigate sample magnetization by standard method of the magnetic field gradual sweeping between two opposite field values to obtain $M(H)$ magnetization loops at different temperatures and for different angles $\alpha$ between the sample and the magnetic field.

\section{Experimental results}

\begin{figure}
\includegraphics[width=\columnwidth]{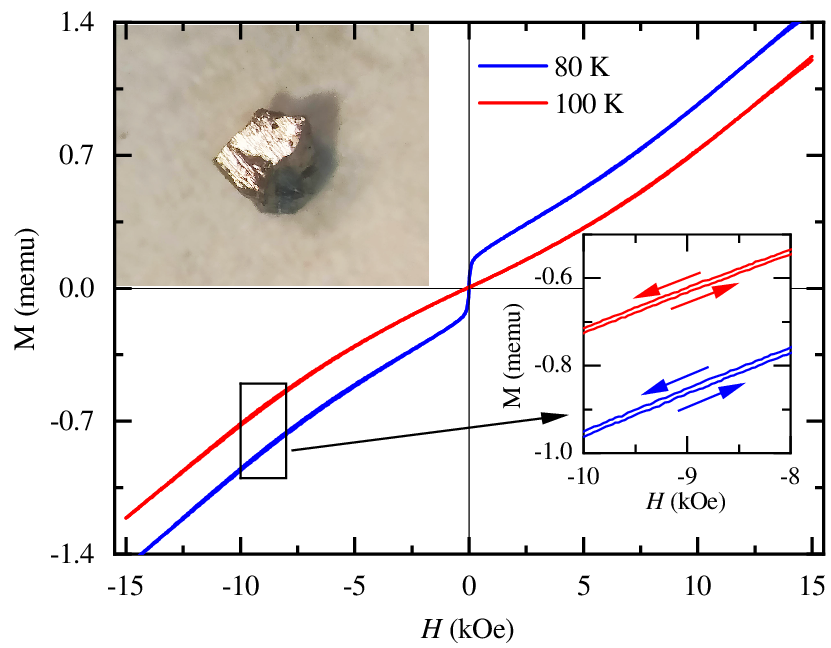}
\caption{(Color online) $M(H)$ magnetization loops for the 5.04~mg   MnTe sample at 80~K and  100~K temperatures (blue and red curves, respectively). $M(H)$ behavior at 100~K well  corresponds to the known antiferromagnetic one for MnTe~\cite{MnTe1,MnTe2,MnTe3,MnTe4,MnTe5,MnTe6}: the sample magnetization is mostly compensated, the small high-field hysteresis (as shown in the right inset) is due to the spin flop processes below the N\'eel vector reorientation field in MnTe~\cite{AHE_MnTe1,AHE_MnTe2}. At 80~K, the $M(H)$ non-linear branches  are shifted vertically, so the well-developed step in $M(H)$ around zero field resembles standard,  ferromagntetic-like behavior. The left inset shows optical image of the mechanically cleaved single crystal MnTe sample. 
 }
\label{fig2}
\end{figure}

Fig.~\ref{fig2} shows $M(H)$ magnetization loops for the 5.04~mg   MnTe sample at 80~K and  100~K temperatures (blue and red curves, respectively). 

Even the 100~K curve shows clearly nonlinear behavior in high magnetic fields, with small hysteresis for the nonlinear branches, see the inset to Fig.~\ref{fig2}.  In contrast, $M(H)$ is nearly linear around zero field. This behavior well corresponds to the known antiferromagnetic one for MnTe~\cite{MnTe1,MnTe2,MnTe3,MnTe4,MnTe5,MnTe6}: in the present field range, we do not reach the N\'eel vector reorientation field, which is between 2 and 3 T for MnTe~\cite{AHE_MnTe1,AHE_MnTe2}, so the sample magnetization is mostly compensated. The small high-field hysteresis due to the spin flop processes above 5~kOe field in Fig.~\ref{fig2}.

\begin{figure}
\includegraphics[width=\columnwidth]{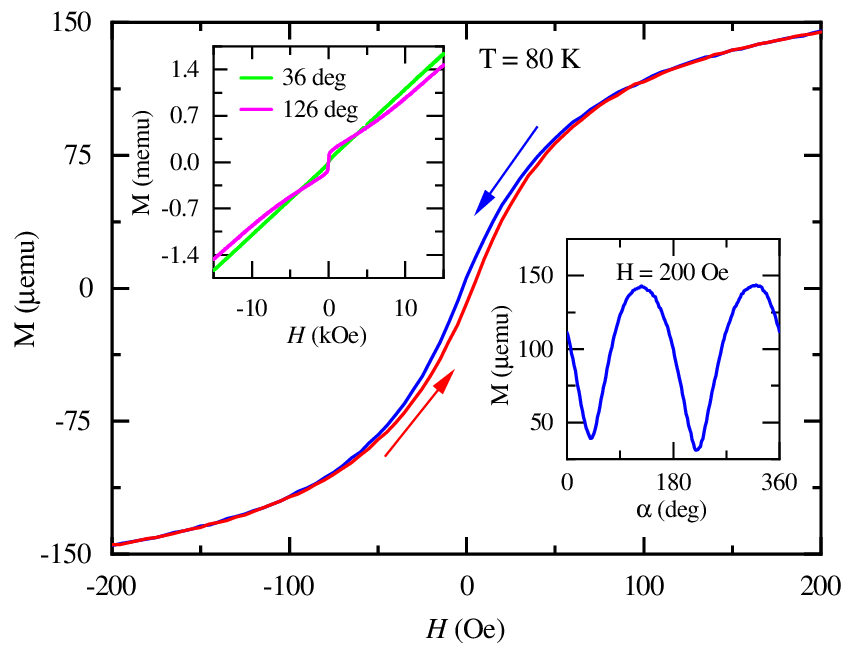}
\caption{(Color online) Low-field hysteresis at 80~K temperature for the 5.04~mg  MnTe sample. The hysteresis saturates at $\pm$100~Oe, standard ferromagnetic behavior  is also confirmed  by the angle dependence  $M(\alpha)$  in the right inset as about 60\%  modulation of the $M(\alpha)$ with 180$^\circ$ periodicity in the magnetic field 200~Oe. The low-field hysteresis is developed around the maxima in  $M(\alpha)$, e.g. the curves in the main field belongs to  $\alpha=126^\circ$. The left inset shows $M(H)$ curves in a wide magnetic field range for two, $\alpha=126^\circ$ and  $\alpha=36^\circ$, angles. In the latter case (the minimum in $M(\alpha)$),  $M(H)$ is linear for the whole magnetic field range. 
 }
\label{fig3}
\end{figure}

At 80~K, the $M(H)$ curves are also consist from two non-linear branches with small high-field hysteresis, as depicted in the inset to Fig.~\ref{fig2}. In addition, the branches are shifted vertically, so there is a well-developed step in $M(H)$ around the zero. This step resembles standard ferromagntetic-like behavior, so we concentrate on the low-field region below, see Fig.~\ref{fig3}.  

Fig.~\ref{fig3} shows narrow but well-defined low-field hysteresis at 80~K temperature. The hysteresis saturates at $\pm$100~Oe, $M(H)$ curves coincide above this value. It is worth to note, that the holder diamagnetic response is about 1~$\mu$emu in this field, which can not affect the presented results.  Standard ferromagnetic behavior  is also confirmed  by the angle dependence  $M(\alpha)$  in the right inset to Fig.~\ref{fig3}: we observe about 60\%  modulation of the $M(\alpha)$ with 180$^\circ$ periodicity in the magnetic field 200~Oe. The low-field hysteresis is observed in the maxima of the $M(\alpha)$ curve, i.e. for $\alpha=126^\circ$ and $\alpha=306^\circ$. In contrast, $M(H)$ is linear for the $M(\alpha)$ minima, as it is shown in the left inset to Fig.~\ref{fig3}  for the whole magnetic field range for for $\alpha=36^\circ$.

\begin{figure}
\includegraphics[width=\columnwidth]{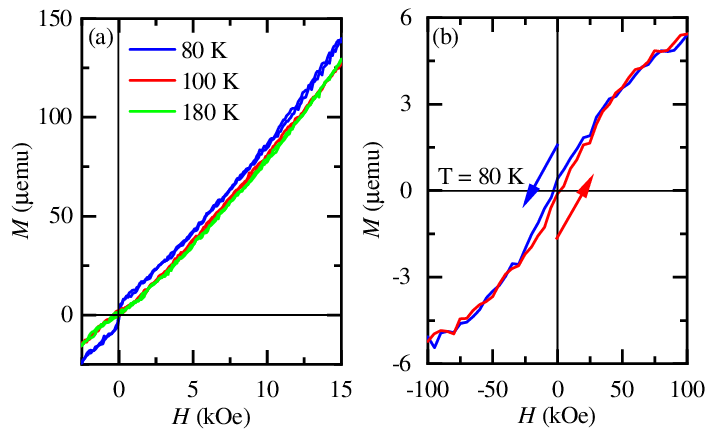}
\caption{(Color online) (a) $M(H)$ curves for the smallest,  0.53~mg sample for different temperatures. All the curves demonstrate small high-field hysteresis, similarly to Fig.~\ref{fig2}. The $M(H)$ curves are nearly coincide at 180~K and 100~K, while there is a well-developed step in $M(H)$ around zero field at 80~K temperature. (b) The low-field hysteresis for this sample around zero field at 80~K temperature. The hysteresis saturates above 100~Oe, the curves are obtained for the maximum of the $M(\alpha)$ angle dependence. 
  }
\label{fig4}
\end{figure}

The reported  $M(H)$ behavior can be qualitatively reproduced for MnTe samples of different sizes. For example, Fig.~\ref{fig4} (a) shows nonlinear $M(H)$ curves for the smallest,  0.53~mg sample for different temperatures. All the curves demonstrate small high-field hysteresis, similarly to Fig.~\ref{fig2}. The $M(H)$ curves are nearly coincide at 180~K and 100~K, while there is a well-developed step in $M(H)$ around zero field at 80~K temperature, which is shown in Fig.~\ref{fig4} (b). The  curves are noisy for the the smallest,  0.53~mg,  sample, however, the $M(H)$ hysteresis can be clearly seen around the zero, it saturates above 100~Oe. It is worth to mention, that the  saturated $M(H)$ value is still one order of magnitude above the holder response in this field range.  

\begin{figure}
\includegraphics[width=\columnwidth]{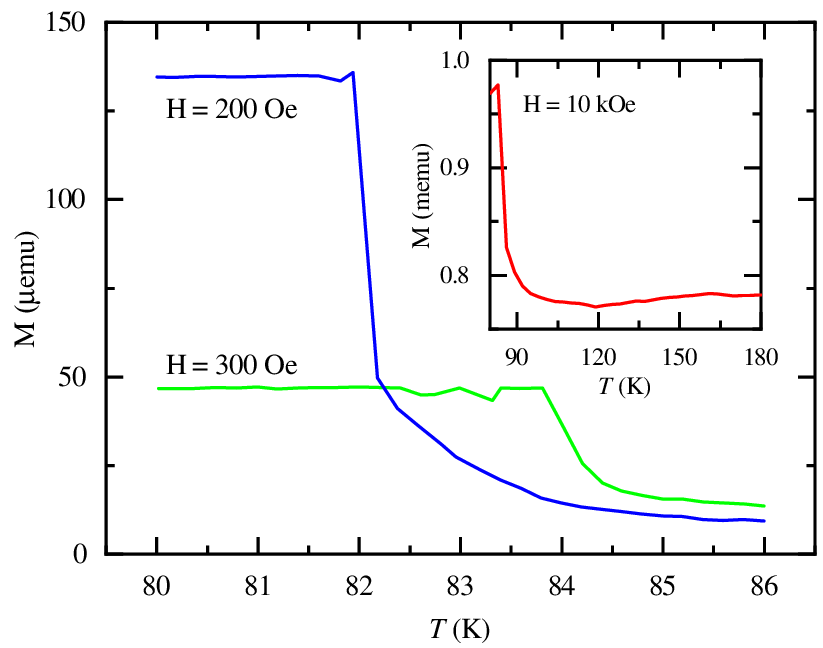}
\caption{(Color online)  $M(T)$ temperature dependence between 80~K and 100~K  in fixed magnetic fields for two, 5.04~mg and 4.65~mg samples, as depicted by blue (200 Oe field) and green (300 Oe)  curves, respectively. The transition temperature shows small variation (82-84~K) from sample to sample. Inset shows $M(T)$ dependence in a wide temperature range at 10~kOe magnetic field. The  curves are obtained for the maximum of the $M(\alpha)$ angle dependence. 
 }
\label{fig5}
\end{figure}

Temperature transition between 80~K and 100~K is shown in Fig.~\ref{fig5} as $M(T)$ curves in fixed magnetic fields for two, 5.04~mg and 4.65~mg samples, see blue and green curves, respectively. $M(T)$ is temperature-independent below the transition, the transition temperature shows small variation (82-84~K) from sample to sample. The $M(T)$ drop is nearly field-independent in Figs.~\ref{fig2} and~\ref{fig4} (a), which is confirmed by $M(T)$ dependence in a wide temperature range in the inset to Fig.~\ref{fig5}. Also, we have checked that the sample holder shows no temperature dependence in this range.

We should conclude, that for temperatures below 85~K the known antiferromagnetic behavior for MnTe~\cite{MnTe1,MnTe2,MnTe3,MnTe4,MnTe5,MnTe6} is accompanied by ferromagnetic one in low fields. 

\begin{figure}
\includegraphics[width=\columnwidth]{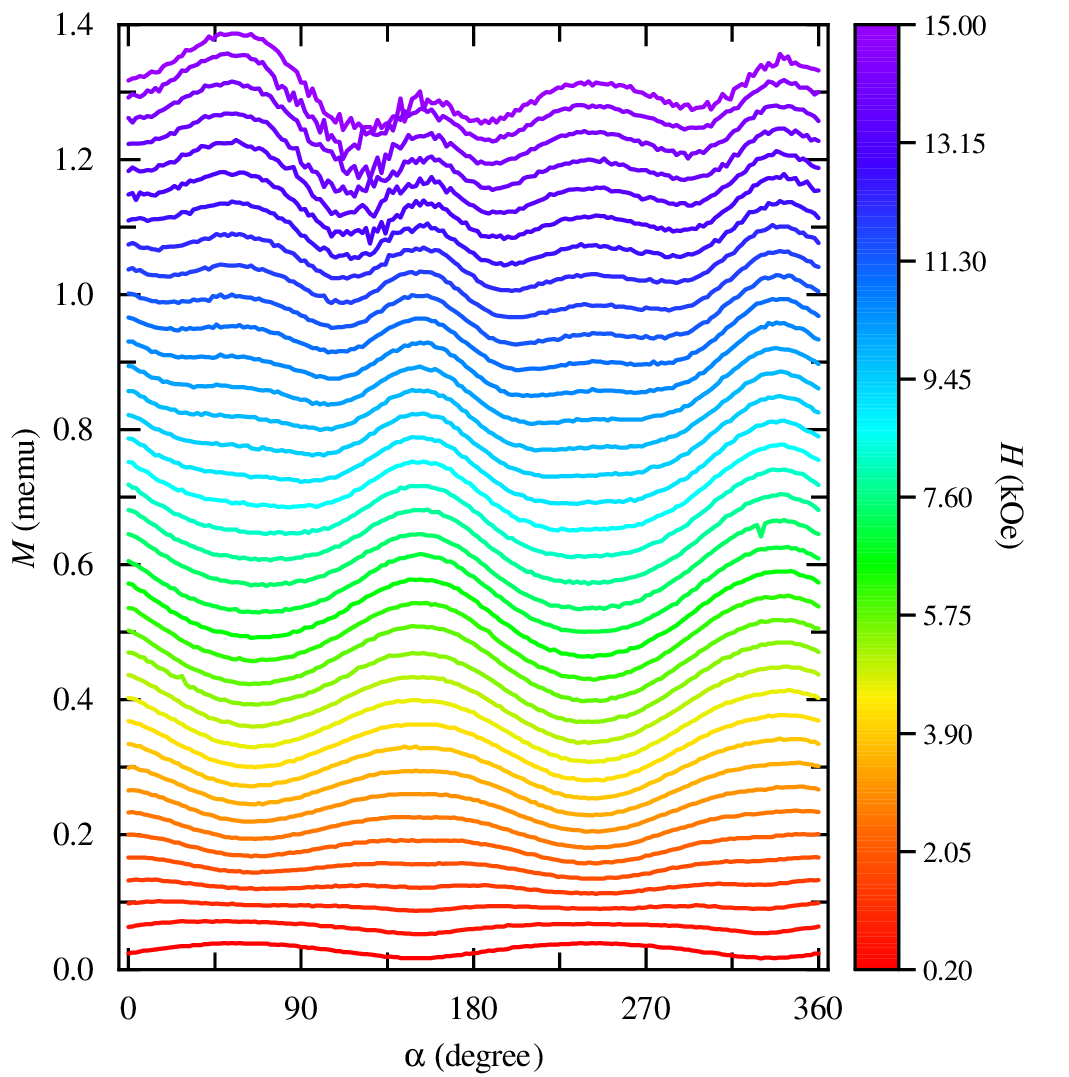}
\caption{(Color online) Crossover from ferromagnetic to antiferromagnetic behavior shown as $M(\alpha)$ curves in different magnetic fields. The curves are presented as obtained, without any additional processing, at 80~K temperature for the 4.65~mg MnTe sample. The mean level reflects the $M(H)$ growth in Fig.~\ref{fig2}, the absolute value of the $M(\alpha)$ modulation is increasing with the magnetic field. In low fields (around 0.2~kOe),  $M(\alpha)$ shows ferromagnetic-like 180$^\circ$ periodicity, as in Fig.~\ref{fig3}. At high magnetic fields (around 15~kOe), the periodicity is changed  to the 90$^\circ$ one. For the intermediate fields (around 6~kOe), one can see interplay between the maxima and the minima in $M(\alpha)$ curves.  This angle dependence is the most striking result of our experiment, while it can not be expected for standard magnetic systems. 
 }
\label{fig6}
\end{figure}

 The crossover from ferromagnetic to antiferromagnetic behavior is shown in Fig.~\ref{fig6} as $M(\alpha)$ curves in different magnetic fields below the transition temperature. In low fields,   $M(\alpha)$ shows ferromagnetic-like 180$^\circ$ periodicity, as it is discussed above. At high magnetic fields, the periodicity is changed  to the 90$^\circ$ one, see e.g. 15~kOe field curve in Fig.~\ref{fig6}. For the intermediate fields, one can see interplay between the maxima and the minima in $M(\alpha)$ curves. In particular, 180$^\circ$ periodic low-field (0.2~kOe) maxima are changed to the still 180$^\circ$ periodic minima around 6~kOe field, which are afterward transformed to the maxima at 15~kOe. In the same time, 180$^\circ$ periodic low-field (0.2~kOe) $M(\alpha)$ minima are monotonously transformed to the  $M(\alpha)$ maxima at 15~kOe field, which gives overall 90$^\circ$ periodicity. This behavior strongly resembles beating patterns in magnetic oscillations~\cite{dorozhkin,beating1,beating2,beating3,beating4}. The $M(\alpha)$ angle dependence is the most striking result of our experiment, while it can not be expected for standard (ferro-, antiferro-, para- and dia-) magnetic systems.
 
\section{Discussion} \label{disc}

The antiferromagnetic ordering with vanishing net magnetization is known~\cite{MnTe1,MnTe2,MnTe3,MnTe4,MnTe5,MnTe6} for $\alpha$--MnTe.  The reorientation field of the N\'eel vector was found to be between 2 and 3~T for MnTe~\cite{AHE_MnTe1,AHE_MnTe2}, so most of investigations were concentrated in high magnetic field range~\cite{AHE_MnTe1,AHE_MnTe2}. 

In high fields, our results well correspond to the known antiferromagnetic behavior, see Fig.~\ref{fig2}: below 15~kOe, we do not reach the reorientation field (2--3~T), so the magnetization of sample is mostly compensated. We observe small high-field hysteresis due to the spin flop processes below 1.5~T. This high-field behavior is nearly independent of temperature, because the N\'eel point is around the room value (307--325~K depending on the thin films or the single crystal samples~\cite{NeelTemp_SC, NeelTemp_film1, NeelTemp_film2}).  

Thus, the high-field behavior for our samples and for the 2.5~$\mu$m thick  MnTe films on SrF2 substrate~\cite{MnTe_magnetization} is qualitatively similar. 
In contrast, the angle dependence in low fields in Fig.~\ref{fig6} does not correlate with the mean-field susceptibility of a collinear uniaxial antiferromagnet for the cases when the field is perpendicular and parallel to the easy axis~\cite{MnTe_magnetization}.


The  low-field region $\pm 200$~Oe was inaccessible in Refs.~\cite{AHE_MnTe1,MnTe_magnetization}, because of dominating diamagnetic contribution from the substrate. Also, the $M(H)$ magnetization loops were not investigated in Ref.~\cite{AHE_MnTe2} for single crystal samples.  In our experiment, the MnTe single crystal is mounted to the sample holder, which contribution is strictly diamagnetic and within 10\% of the measured value, see the Samples section for details. Thus, our experimental setup allows high-resolution measurements in low fields, in contrast to previous investigations.  

For our samples, the MnTe composition was verified by energy-dispersive X-ray spectroscopy and the powder X-ray diffraction analysis, so the low-field hysteresis can not appear from incorrect stoichiometry or Mn oxides~\cite{impur1,impur2}. Moreother, the samples show  an abrupt drop of the  magnetization around 82-84~K, which we should identified as the sign of the first-order magnetic transition~\cite{mag_first_order1,mag_first_order2}. On the other hand, the magnetic susceptibility drop has been reported for MnTe around 80~K in zero magnetic field, which is not accompanied by any structural transition~\cite{Xi_step,Xi_step1}. Thus, our low-field magnetization results well correspond to the known MnTe properties, so they are not defined by any sample disadvantages. 

Altermagnetic candidate MnTe is expected to have zero nonrelativistic net magnetization~\cite{alter_mazin,MnTe_Mazin}. On the other hand, it is accepted, that the principle origin of AHE~\cite{AHE_MnTe1,AHE_MnTe2} in MnTe, and, therefore  of weak remanent  magnetization~\cite{AHE_MnTe2} is the spin-orbit coupling in the valence orbitals~\cite{Dichroism}.  Thus, we should attribute the observed low-field $M(H)$ hysteresis to the effects of spin-orbit coupling. 

The effects of spin-orbit coupling in this material has been previously investigated by temperature-dependent angle-resolved photoelectron spectroscopy and by disordered local moment calculations~\cite{MnTe_SO}. There were the emergence of a relativistic valence band splitting concurrent with the establishment of magnetic order. It seems to be important, that the  observed splitting is well-resolved only below 100~K in Ref.~~\cite{MnTe_SO}, which is consistent with our experiment. 

In this case,  the  interplay between maxima in low fields and minima in high magnetic fields in  $M(\alpha)$ angle dependence in Fig.~\ref{fig6} is the standard beating pattern~\cite{dorozhkin,beating1,beating2,beating3,beating4}, where the behavior in low fields is determined by spin-orbit coupling, while the spin flop processes are dominating in high magnetic field. The spin-orbit splitting can be suppressed by temperature, which is reflected as an abrupt drop of the  magnetization around 82-84~K in Fig.~\ref{fig5}. Thus, in contrast to the charge transport, our experiment directly shows the effect of weak spin-orbit coupling in MnTe altermagnet candidate, with crossover from relativistic to non-relativistic net magnetization while increasing the magnetic field.

It is well known, that in altermagnets,  spin magnetic moments are fully compensated when spin-orbit coupling is zero, but nonzero SOC, enabling coupling between spins and alternating local structures, can result in a nonzero net magnetic moment~\cite{alter_ferro}. In other words, symmetry allows ferromagnetic behaviors (nonzero net magnetic moment, AHE)  only due to the  spin-orbit coupling, which we confirm in our experiment.

\section{Conclusion}
As a conclusion, we experimentally study magnetization reversal curves for MnTe single crystals, which is the altermagnetic candidate. Above 85~K temperature, we confirm the antiferromagnetic  behavior of magnetization $M$, which is known for $\alpha$--MnTe. Below 85~K, we observe anomalous low-field magnetization behavior, which is accompanied by the sophisticated  $M(\alpha)$ angle dependence with beating pattern as the  interplay between $M(\alpha)$ maxima and minima: in low fields,   $M(\alpha)$ shows ferromagnetic-like 180$^\circ$ periodicity, while at high magnetic fields, the periodicity is changed  to the 90$^\circ$ one. This angle dependence is the most striking result of our experiment, while it can not be expected for standard magnetic systems. In contrast, in altermagnets,  symmetry allows ferromagnetic behavior only due to the  spin-orbit coupling. Thus, we claim that our experiment shows the effect of weak spin-orbit coupling in MnTe, with crossover from relativistic to non-relativistic net magnetization, and, therefore, we experimentally confirm altermagnetism in MnTe.

\acknowledgments

We wish to thank S.S~Khasanov for X-ray sample characterization and Vladimir Zyuzin for valuable discussions.  We gratefully acknowledge financial support  by the  Russian Science Foundation, project RSF-24-22-00060, https://rscf.ru/project/24-22-00060/

\end{document}